\documentclass[twocolumn,aps,pre,showpacs]{revtex4}
\usepackage{graphicx}
\usepackage{latexsym}
\usepackage{amssymb}
\usepackage{amsmath}
\begin{document}
\def\d{{\rm d}}
\def\ex{{\rm e}}
\def\U{{\bf U}}
\def\x{{\bf x}}
\def\A{{\bf A}}
\def\W{{\bf W}}
\def\k{{\rm k}}
\def\R{{\bf R}}
\def\p{{\bf p}}
\def\E{{\bf E}}
\def\bOmega{{\boldsymbol{\Omega}}}
\def\bsigma{{\boldsymbol{\sigma}}}
\def\smalk{{\scriptscriptstyle{\rm k}}}
\def\beq{\begin{equation}}
\def\eeq{\end{equation}}
\font\brm=cmr10 at 24truept
\font\bfm=cmbx10 at 15truept
\title{Orientation dynamics of weakly Brownian \\ particles in periodic viscous flows}
\author{Piero Olla}
\affiliation{ISAC-CNR, Sez. Lecce,  I--73100 Lecce, Italy.}
\date{\today}

\begin{abstract}
Evolution equations for the orientation distribution of axisymmetric
particles in periodic flows are derived in the regime of small
but non-zero Brownian rotations. The equations are based on a multiple
time scale approach that allows fast computation of the relaxation
processes leading to statistical equilibrium. The approach has been 
applied to the calculation of the effective viscosity of a thin disk
suspension in an oscillating strain flow.
\end{abstract}

 \maketitle

\section{INTRODUCTION}

The rheological properties of a suspension will depend, when
the particles are non-spherical, on the
orientation taken by the particles in response to the external flow. For a few particle 
shapes (e.g. the case of the ellipsoid \cite{jeffery22}), 
equations for the rotation dynamics exist in closed form,
and it is possible to determine the orientation distribution of
the particles in suspension. However, unless a
mechanism for the achievement of a statistical equilibrium is
introduced, the orientation distribution will depend on the
state in which the suspension is prepared initially.
In the case of microscopic particles, one such mechanism 
is provided by Brownian rotations \cite{taylor23}.  
It is still unclear
whether inertia and interaction with other particles may contribute
to the equilibration mechanism.

An equilibrium distribution could be achieved alternatively by the presence of 
chaos in the rotation dynamics; unfortunately, the importance of 
chaos turns out to be small in most situations.
In the case of a simple shear and axisymmetric particles, the particle motion 
is periodic \cite{jeffery22}. This motion becomes aperiodic in the case of a time-dependent flow,
but remains non-chaotic for axisymmetric particles \cite{szeri92}.
Chaos arises in the motion of a triaxial ellipsoids in 
a simple shear \cite{yarin97}, but, depending on the axis ratios, large domains of initial 
conditions remain associated with regular orbits and to the absence of a uniquely
defined equilibrium distribution. Furthermore, for weak Brownian motion, the
regular regions will act as attractors for the chaotic orbits and will provide the 
bulk of the orientation distribution. 

The equilibrium orientation distribution of a Brownian particle has been determined
in various important limit regimes, depending on the value of the Peclet number $Pe$, 
defined as the ratio of the velocity gradient and the angular diffusivity
(which has the dimension of a frequency).
The case of strong Brownian rotation was considered by Burgers \cite{burgers38}
leading to an orientation distribution that in first approximation can be 
considered isotropic. A systematic perturbation theory in powers of $Pe$ was
introduced in \cite{peterlin38}, allowing the calculation of the effective 
viscosity of dilute suspensions, for values of $Pe$ up to $20-30$ \cite{scheraga55}.

More interesting is the case of weak Brownian motion, in which the form of the
equilibrium distribution is determined by the structure of the orbits in orientation 
space, which in turn depends on the imposed flow. 
A technique for the determination of the equilibrium distribution of weakly 
Brownian particles, based 
on singular perturbation analysis of the diffusion equation in orientation space,
was derived in \cite{leal72} for the case of axisysmmetric particles in a simple
shear. 

In the present paper, an alternative approach will be presented, 
based on the perturbative determination of the orbits
in orientation space. This approach will 
appear to be appropriate in the case the flow is time-dependent, and, more in 
general, when analytical expressions for the unperturbed orbits are not available.
For the sake of definiteness, the dynamics of a small disk
in the field of an oscillating strain flow will be analyzed, and its contribution 
to the medium effective viscosity determined. 
This kind of flow could be obtained by means of a four-roll mill, as described in 
\cite{szeri92}; more interestingly, as it will be illustrated in the next section,
an oscillating pure strain is what is seen by a particle in the velocity field 
of a gravity wave. This turns out to have application to models of wave 
propagation in polar seas. In fact, under cold and windy conditions,
high concentrations of millimeter size ice crystals, called frazil ice,
are generated in the water, modifying the medium viscosity and leading
to increased wave damping
\cite{martin81,newyear97}. (Given the crystal size, $Pe$ is in this 
case typically very large).

This paper is organized as follows. In the next section the orientation 
dynamics of a small disk will be 
analyzed in the absence of Brownian rotations. In Section III, the
diffusion and drift across orbits in orientation space, produced by
a weak noise, will be calculated perturbatively. In Section IV 
the results will be applied to the calculation of the effective
viscosity of a dilute small disk suspension.
Section V is devoted to conclusions.

\section{UNPERTURBED ORIENTATION DYNAMICS}
Consider the motion of a particle in a velocity field $\U=(U_1,U_2,0)$, which, 
at the particle position, has zero vorticity, and strain ${\bf E}=[\nabla\U+(\nabla\U)^{\rm T}]/2$
with components in the directions $x_1$ and $x_2$:
\beq
{\bf E}
=e\left( 
\begin{array}{rr} 
(1+\alpha)\cos\omega t, &   (1-\alpha)\sin\omega t  \\
(1-\alpha)\sin\omega t, & -(1+\alpha)\cos\omega t   
\end{array} 
\right).
\label{1}
\eeq
The interest in Eq. (\ref{1}) is that it describes what is experienced by 
a fluid element in a gravity wave. In fact, the velocity field of a 
small amplitude gravity wave, would read,
choosing the $x_2$ axis pointing downward from the unperturbed 
water surface at $x_2=0$ \cite{newman}:
\beq
\begin{array}{rr}
U_1 = \widetilde{U}[\ex^{-kx_2}+\ex^{k(x_2-2h)}]\sin (kx_1-\omega t),
\\
U_2= \widetilde{U}[\ex^{-kx_2}-\ex^{k(x_2-2h)}] \cos (kx_1-\omega t),
\end{array}
\label{4}
\eeq
where $h$ is the water depth. From the potential nature of the flow, the vorticity
of the flow $\bOmega=[\nabla\U-(\nabla\U)^{\rm T}]/2$
is zero and the strain at the 
sea surface $x_2=0$, has the form given by Eq. (\ref{1}), with $e=k\widetilde{U}$,
$\alpha=\exp(-2kh)$, and an initial phase different from zero for $x_1\ne 0$. For small
wave amplitudes, the displacement of a particle initially at the water surface will 
be small and the strain field experienced ${\bf E}(\x(t),t)\simeq{\bf E}(\x(0),t)$
will be given by Eq. (\ref{1}).  Expressions
for the strain field in the form of Eq. (\ref{1}) can be shown to occur also for $h\to\infty$,
from the
superposition of progressive and regressive waves, with $e=k\widetilde{U}^+
>k\widetilde{U}^-$, $\alpha=\widetilde{U}^-/\widetilde{U}^+$, and $\widetilde{U}^\pm$
the amplitudes of the two wave components.

For $\alpha=0$, Eq. (\ref{1}) describes a constant strain rotating with frequency
$\omega/2$ around the $x_3$-axis, which, in the the gravity wave example,
is associated with particle orbits that are perfectly circular \cite{newman}.
Transforming to the rotating reference frame
leads to the new expression for the strain field
\beq 
{\bf E}=e 
\left(
\begin{array}{rr}
\alpha\sin 2\omega t, & 1+\alpha\cos 2\omega t\\
1+\alpha\cos 2\omega t, &-\alpha\sin 2\omega t  
\end{array}
\right)
\label{2} 
\eeq
(the initial phase of the rotation has been chosen
to produce, at $t=0$, a strain field with expanding 
direction at $\pi/4$ with respect to the new $x_1$ axis; see Appendix A).
In the rotating reference frame, an additional vorticity field is produced:
\beq
\bOmega= \frac{\omega}{2}
\left(\begin{array}{rr}
        0 &    1 \\
        -1 &   0 \\
\end{array}\right).
\label{3}
\eeq
For $\alpha\ne 0$, this is a time-dependent planar flow, of the kind discussed in \cite{szeri92},
which is known to produce aperiodic behaviors in the particle orientation dynamics.

\begin{figure}
\begin{center}
\includegraphics[draft=false,width=4.5cm]{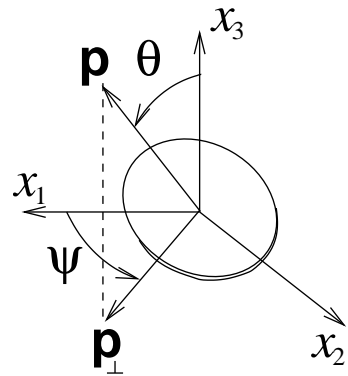}
\caption{
The coordinate system. The axes $x_i$ are in the rotating reference frame.
}
\label{shallfig1}
\end{center}
\end{figure}

The motion of a revolution ellipsoid, with symmetry axis identified by the versor $\p$, 
in the presence of the strain and vorticity fields $\E$ and $\bOmega$, 
is described by the Jeffery's equations \cite{jeffery22}:
\beq
\dot\p =\bOmega\cdot\p+G [\E\cdot\p - (\p\cdot\E\cdot\p)\p].
\label{5}
\eeq
The parameter $G$ gives the ellipsoid eccentricity, defined in terms of the particle aspect ratio
$r=a/b$, where $a$ and $b$ are respectively along and perpendicular to the symmetry axis,
by means of the relation
$$
G=\frac{r^{2}-1}{r^{2}+1}\ .
$$
Introducing polar coordinates (see Fig. \ref{shallfig1}) and 
normalizing time and vorticity by the strain strength $e$:
$\omega\to\omega/(-2Ge)$ and $t\to -Get$, the Jeffery's equation (\ref{5}), 
using Eqs. (\ref{2}) and (\ref{3}), will take the form:
\beq
\left\{ \begin{array}{l}
\dot\psi = -\omega+\beta(\psi,t),
 \\
\dot c = -\frac{1}{2}\beta'(\psi,t)c,
\end{array}\right.
\label{6}
\eeq
with $c=\tan\theta$,
dot and prime indicating respectively $\d/\d t$ and $\partial/\partial_\psi$, 
and:
\beq
\left\{ \begin{array}{l}
\beta(\psi,t)=-\cos 2\psi-\alpha\cos(4\omega t+2\psi),\qquad
\\
\beta'(\psi,t)=2[\sin 2\psi+\alpha\sin(4\omega t+2\psi)]
\end{array}\right.
\label{7}
\eeq
(more details in Appendix A).

Following \cite{szeri92}, the orbits can be classified studying the Poincare map
$P_n(\bar\psi)={\rm mod}(\psi(nT$ $|\bar\psi),\pi)$, with $T=\frac{\pi}{2\omega}$ 
the period of $\beta$, where $\psi(t|\bar\psi)$ obeys the first of Eq. (\ref{6}) 
with $\psi(0|\bar\psi)=\bar\psi$. This eliminates the explicit time dependence 
from the dynamics, and will allow to isolate
the slow, noise produced deviation between orbits, from the fast motion 
along them (see next sections).

\begin{figure}
\begin{center}
\includegraphics[draft=false,width=8.5cm]{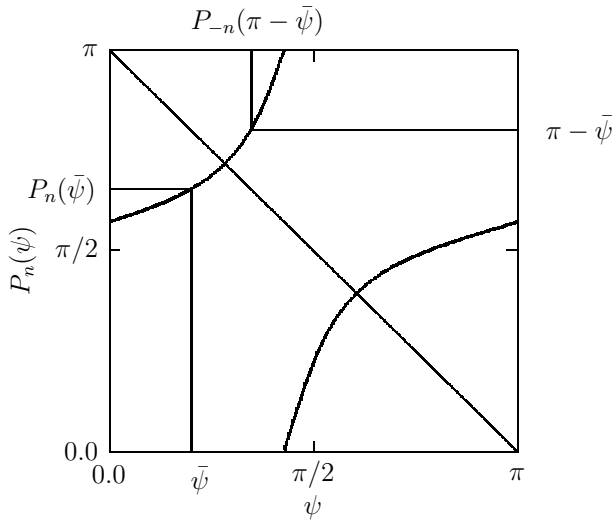}
\caption{
Symmetry of the Poincare map for the dynamics of Eqs. (\ref{6},\ref{7}). Values of the parameters:
$\omega=1.4$, $\alpha\simeq 0.37$, $n=1$; $G<0$ (oblate ellipsoid). From Eq. (\ref{8}), one has
that $P_{-n}(\pi-\psi)=\pi-P_n(\psi)$ and therefore the plot is symmetric under reflection 
across the diagonal line $P_n=\pi-\psi$.
}
\label{shallfig2}
\end{center}
\end{figure}

\begin{figure*}
\begin{center}
\includegraphics[draft=false,width=14cm]{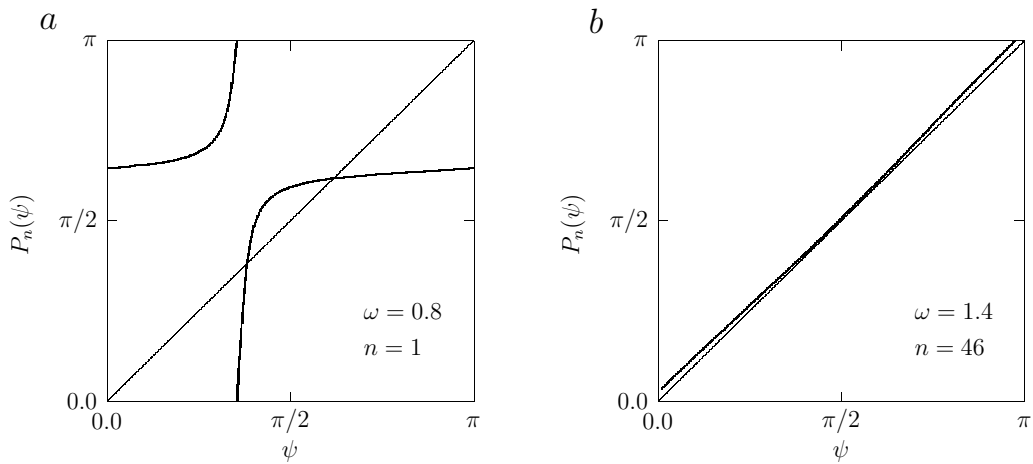}
\caption{
Poincare map for the coherent orientation ($a$) and the random orientation regime ($b$) of
an oblate ellipsoid in a shallow water wave with $\alpha\simeq 0.37$. Notice in case ($a$)
the stable fixed point at
$\psi>\pi/2$ and the unstable one at $\psi<\pi/2$. With prolate ellipsoids, the
fixed points would have been exchanged. The value of $n$ in the random 
orientation case ($b$) has been chosen to lead to approximately closed orbits. Notice that
$\psi=\pi/2$ remains the best approximation to a fixed point (i.e. a closed orbit
of period $nT$).
}
\label{shallfig3}
\end{center}
\end{figure*}

Some properties of the Poincare map can be
obtained from inspection of Eqs. (\ref{6}-\ref{7}). In particular, it is possible 
to see, from $\beta(\psi,t)=\beta(-\psi,-t)$ and the form of Eq. (\ref{6}), 
that the following relation holds:
\beq
P_{-n}(-\bar\psi)=-P_n(\bar\psi),
\label{8}
\eeq
and therefore the Poincare map is symmetric under 
the double reflection $\{\psi,P_n\}\to\{\pi-\psi,\pi-P_n\}$ (see Fig. \ref{shallfig2}).
This has the consequence that fixed points, when present, would come in pairs 
located symmetrically around $\psi=\pi/2$ (see Fig. \ref{shallfig3}a).

A fixed point in the Poincare map will be associated with a periodic $\psi$ and
correspond to coherent orientation of the particles. This regime is clearly
produced by the aligning effect of strain on the particle dynamics, which 
becomes dominant in the small $\omega$ range. In the present case, Eqs. 
(\ref{6}-\ref{7}) appear to lead at most to a pair of fixed points, of 
which the the stable one is located at $\psi>\pi/2$ (see Fig. \ref{shallfig3}a).
The stable fixed point tends to $\psi=3\pi/4$ at $\omega=0$, corresponding to alignment
of the long ellipsoid axis with the strain expanding direction.

A transition to a coherent
orientation regime is predicted in the case of a deep 
gravity wave \cite{decarolis05} at the crossover frequency $\omega_c=1$ 
(no superposition of regressive and progressive components).
Now, at $\omega_c\simeq 1$, the small wave amplitude approximation,
leading to Eqs. (\ref{6}-\ref{7}) ceases to be valid (back to dimensional 
units, one would have for the particle displacement
$\Delta x\sim\widetilde{U}/\omega_c\sim k^{-1}$). Nonetheless, 
the linearized theory provides a qualitatively correct
picture, as a transition to a coherent orientation regime is observed experimentally
in wave tanks, although with a larger transition frequency $\omega_c\simeq 1.43$
\cite{martin81}.
The presence of a coherent orientation regime appears to be preserved for
$\alpha\ne 0$ with a crossover frequency $\omega_c$ slowly decreasing as 
$\alpha\to 1$ (at $\alpha\simeq 0.82$, one has still $\omega_c\simeq 0.7$).
The decrease in the crossover frequency can be explained 
in terms of the destabilization of the fixed particle orientation in the rotating frame, 
by the oscillating strain component of Eq. (\ref{2}).

The alternative regime of random particle orientation, is 
associated with $|\psi(nT)|$ increasing monotonously with $n$, with $P_n(\psi)$
generally aperiodic. In this case, from continuity of $\psi(t|\bar\psi)$, $P_n(\bar\psi)$
will be topologically equivalent to an irrational rotation, and the sequence $P_n(\bar\psi)$
originating from a single $\bar\psi$ will fill densely the interval $[0,\pi]$.
An ergodic property is then satisfied, i.e. it is possible to calculate averages
over $\psi$ as time averages. Furthermore, from Poincare recurrence, the $P_n(\bar\psi)$
sequence will come arbitrarily close to the initial condition for some $n$. 

An important property is the following: if the orbit starting from a certain $\bar\psi$
is approximately closed at $t=nT$, as shown in Fig. \ref{shallfig3}b, the same
will occur with the orbits starting from any other initial condition. 
In fact, if $P_n(\bar\psi)-\bar\psi$ is small, 
the same will be true also for $P_n(P_m(\bar\psi))-P_m(\bar\psi)$ with $m$ arbitrary. 
This is consequence of the dynamics of $\psi$ being not chaotic
(i.e. neighbouring trajectories do not separate asymptotically).
Hence, exploiting the fact that, from ergodicity, $P_m(\bar\psi)$ fills densely 
the whole interval $[0,\pi]$, $P_n(\psi)-\psi$ will be small for $\psi$ generic.

Turning to the polar angle, if the orbit in $\psi$ is approximately closed at $nT$, also
$c(nT|\bar c,\bar\psi)$ will come arbitrarily close to the initial condition 
$c(0|\bar c,\bar\psi)=\bar c$. 
Hence, to identify approximately closed orbits, it is sufficient to look for
recurrence of the Poincare map $P_n(\bar\psi)$.
To see why this property holds, the two of Eq. (\ref{6}) can be integrated to give:
$$
\ln\frac{c(nT|\bar c,\bar\psi)}{\bar c}=-\frac{1}{2}
\ln\frac{\partial P_n(\bar\psi)}{\partial\bar\psi},
$$
and the condition $c(nT|\bar c,\bar\psi)\simeq \bar c$ will be satisfied provided
$\partial P_n/\partial\bar\psi\simeq 1$. In the present case, the condition 
$\partial P_n/\partial\bar\psi\simeq 1$ is satisfied provided 
$P_n(\bar\psi)\simeq\bar\psi$, i.e. if the orbit is approximately closed.
The contrary would require $\partial P_n/\partial\bar\psi$ to oscillate in $\bar\psi$
around $\partial P_n/\partial\bar\psi=1$. However, if $|P_n(\bar\psi)-\bar\psi|<\epsilon$ 
for for some small $\epsilon$, the difference
$\partial P_n/\partial\bar\psi-1$ could remain of $O(1)$ in intervals at most 
of length $O(\epsilon)$, in which one would have in consequence
$\partial^2 P_n/\partial\bar\psi^2=O(\epsilon^{-1})$. But this is prevented
from smothness of the trajectories and of the functions $\beta$ and $\beta'$.

\section{THE EFFECT OF NOISE}
\subsection{Coherent orientation regime}
Noise produces qualitatively different effects in the coherent and in the random 
orientation regimes. In the coherent orientation regime, the main effect
is smearing the transition to the random orientation regime. It is easier
to describe what happens at the transition for $\alpha=0$, where
analytical expressions for $\psi(t)$ and $c(t)$ are available. When the crossover
frequency $\omega_c=1$ is approached from above, i.e. from the random orientation
regime, the rotation period  for $\psi$: $T_r$, in the absence of noise, will tend to 
infinity like $(\omega^2-1)^{-\frac{1}{2}}$ \cite{decarolis05}. For $\omega<1$,
the particle is stuck at the stable fixed point $\bar\psi=(\cos^{-1}\omega+\pi)/2$ 
and the period is by definition infinite. It turns out, that adding a small noise eliminates
divergence of $T_r$ for $\omega\to\omega_c$, as noise allows
the particle to escape from the fixed point.

The time of escape
from the fixed point can be expressed in terms of the inverse of the probability 
for $\psi$ to reach the border of the basin of attraction for $\bar\psi$, which
is $\psi=\pi/2$;
in other words: $T_r^{-1}\sim P(\psi<\pi/2)$.
The probability $P(\psi<\pi/2)$
could be roughly estimated, approximating the dynamics of $\psi$ by the one of the
Langevin equation
obtained through linearization around $\bar\psi$, in the presence of noise, 
of the first of Eq. (\ref{6}):
$$
\d\psi=2(\psi-\bar\psi)\sin 2\bar\psi\ \d t+D^\frac{1}{2}\,\d W.
$$
Here $\d W$ is the Brownian noise increment (Wiener process \cite{gardiner}): 
$\langle\d W\rangle=0$, 
$\langle\d W^2\rangle=\d t$, and $D$ is 
supposed small. This Langevin equation leads to the PDF (probability density function) 
for $\psi$ \cite{gardiner}:
$$
\rho(\psi)={\rm const.}\ \exp\Big(-\frac{2|\sin 2\bar\psi|}{D}(\psi-\bar\psi)^2\Big).
$$
For small values of $D$, $P(\psi<\pi/2)\sim\rho(\pi/2)$ and therefore:
$$
T_r\sim\exp\Big(\frac{2|\sin 2\bar\psi|}{D}(\bar\psi-\pi/2)^2\Big).
$$
Thus, the rotation period is exponentially long in the inverse noise amplitude
and the effect of Brownian rotations on the orientation distribution, which
is governed by the stable fixed point of $P_n(\psi)$, will vanish in the zero 
noise limit.

\subsection{Random orientation regime}
In the random orientation regime, the role of noise 
in determining an equilibrium orientation distribution is fundamental.
If Brownian rotations were
strictly zero, the evolution of the PDF $\rho(\psi,c;t)$ would be given
by propagation along the unperturbed trajectories described by Eq. (\ref{6}),
which, from now on, will be identified by subscript zero: 
\beq
\rho(\psi_0(t|\bar\psi),c_0(t|\bar c,\bar\psi);t)=\rho(\bar\psi,\bar c;0)J(\bar\psi,\bar c),
\label{8.5}
\eeq
where
$J(\bar\psi,\bar c)=|\det[(\d\psi_0,\d c_0)/(\d\bar\psi,\d\bar c)]|^{-1}$ is
the Jacobian of the transformation $\{\bar\psi,\bar c\}\to\{\psi^0,c^0\}$.
This PDF is itself recurrent at the recurrence times $t_i=n_iT$, $i=1,2,...$, for which 
$P_{n_i}^0(\bar\psi)\simeq\bar\psi$, and therefore also
$c_0(t_i|\bar c,\bar\psi)\simeq\bar c$ (see the end of last section).
Hence, memory of any initial PDF $\rho(c,\psi;0)$ would be preserved at 
arbitrary large $t_i$:
$\rho(c,\psi;t_i)\simeq\rho(c,\psi;0)$ and no relaxation to equilibrium
would be possible.

Adding noise allows to reach statistical equilibrium in a time of
the order of the inverse of the noise amplitude. Restricting to the
discrete times $nT$, to make the process stationary, the equilibrium
PDF for $\psi$ is obtained from ergodicity and is the unique stationary PDF
$\rho_E(\psi)={\rm const}.\,|\omega+\beta(\psi,0)|^{-1}$. The statistics of
$c$, can then be described in terms of the conditional PDF
$\rho(c|\psi)=\rho_E(\psi,c)/\rho_E(\psi)$, where $\rho_E(\psi,c)$ is the equilibrium
joint PDF at the instants $t=nT$.
Notice that, from ergodicity of $\psi$, it is sufficient to prescribe the form of 
$\rho(\bar c|\bar\psi)$ at a single position $\bar\psi$. In fact, to obtain
$\rho(c|\psi)$ at $\psi\ne\bar\psi$, to lowest order in the noise, it is
sufficient to propagate Eq. (\ref{8.5}) to $t=nT$
and exploit the fact that $P_n^0(\bar\psi)=
{\rm mod}(\psi_0(nT|\bar\psi),\pi)$ is dense in $[0,\pi]$.

The first step to obtain a kinetic equation for $\rho(\bar c|\bar\psi)$, is to calculate 
the noise produced trajectory separation 
$\{c(t|\bar c,\bar\psi)-c_0(t|\bar c,\bar\psi),\psi(t|\bar\psi)-\psi_0(t|\bar\psi)\}$ at the
recurrence times $t=t_i$, where, choosing appropriately $t_i$, the differences
$c_0(t_i|\bar c,\bar\psi)-\bar c$ and $P^0_{n_i}(\bar\psi)-\bar\psi$ 
can be made small at pleasure.  However, 
since the conditioning in $\rho(\bar c|\bar\psi)$ is at $\bar\psi$ and not at 
$\psi(t|\bar\psi)$, it is then necessary to correct the first step and
calculate the
deviation $c(t|\bar c,\bar\psi)-c_0(t|\bar c,\bar\psi)$ at the time $\hat t_i$
(equal to $t_i$ only in the zero noise limit),
at which mod$(\psi(\hat t_i|\bar\psi),\pi)$ and $\bar\psi$ are strictly equal.
This means considering, instead of a Poincare map synchronized with the period of $\beta$,
the one synchronized with the rotation period in $\psi$, i.e. with the crossing of $\bar\psi$ 
by $\psi$.

Accounting for the effect of Brownian rotations, the Jeffery's equations will read (see
Appendix B):
\beq
\left\{ \begin{array}{l} 
\d\psi=[-\omega+\beta(\psi,t)]\d t+D^{1/2}g^\frac{1}{2}(c)\d W_\psi,
\\
\d c=[-\frac{1}{2}\beta'(\psi,t)c+Df(c)]\d t+D^\frac{1}{2}h^\frac{1}{2}(c)\d W_c,
\end{array}\right.
\label{9}
\eeq
where $D$ has the meaning of a diffusion constant (in the present dimensionless units,
$D^{-1}=Pe$), 
$\d W_k$, with $k=\psi,c$, are the Brownian increments:
\beq
\langle\d W_k\rangle=0,
\qquad
\langle\d W_k\d W_j\rangle=\delta_{kj}\d t,
\label{10}
\eeq
and the functions $f$, $g$ and $h$ are given by (see again Appendix B):
\begin{eqnarray}
f(c)=\frac{1}{c}(1+c^2)(\frac{1}{2}+c^2),
\nonumber
\\
g(c)=\frac{1}{c^2}+1
\quad
{\rm and}
\quad
h(c)=(1+c^2)^2.
\label{11}
\end{eqnarray}
The unperturbed orbits, as already mentioned, indicated by
$\{\psi_0,c_0\}$, obey Eq. (\ref{6}):
\beq
\left\{ \begin{array}{l}
\dot\psi_0 = -\omega+\beta_0,
\\
\dot c_0 = -\frac{1}{2}\beta'_0c_0,
\end{array}\right.
\label{12}
\eeq
with $\beta_0=\beta(\psi_0,t)$ and similar definition for $\beta'_0$.
For small noise, the correction can be determined as an expansions in powers of $D^{1/2}$:
$\psi=\psi_0+\psi_{1/2}+\psi_1+...$ and similarly for $c$, with the initial condition
$\psi_k(0)=c_k(0)=0$ for $k>0$. The lowest order correction is obtained from 
linearization of Eq. (\ref{9}) around $\{\psi_0,c_0\}$:
\beq
\left\{ \begin{array}{l} 
\d\psi_{1/2}=\beta'_0\psi_{1/2}\d t+D^{1/2}g_0^\frac{1}{2}\d W_\psi,
\\
\d c_{1/2}=[2\beta_0c_0\psi_{1/2}-\frac{1}{2}\beta'_0c_{1/2}]\d t
+D^\frac{1}{2}h^\frac{1}{2}_0\d W_c.
\end{array}\right.
\label{13}
\eeq
From Eq. (\ref{10}) and from linearity of Eq. (\ref{13})
$\langle\psi_{1/2}\rangle=\langle c_{1/2}\rangle=0$, but $\psi_{1/2}$ and $c_{1/2}$ are not
zero. The covariance equations obtained from Eq. (\ref{13}) are:
\beq
\begin{cases}
\frac{\d}{\d t}\langle\psi_{1/2}^2\rangle=2\beta'_0\langle\psi_{1/2}^2\rangle+Dg_0,
\\
\frac{\d}{\d t}\langle\psi_{1/2}c_{1/2}\rangle
=\frac{1}{2}\beta'_0\langle\psi_{1/2}c_{1/2}\rangle+2\beta_0c_0\langle\psi_{1/2}^2\rangle,
\\
\frac{\d}{\d t}\langle c_{1/2}^2\rangle=4\beta_0c_0\langle\psi_{1/2}c_{1/2}\rangle
-\beta'_0\langle c_{1/2}^2\rangle+Dh_0,
\end{cases}
\label{14}
\eeq
and lead to a diffusion contribution to the deviation.
To obtain the drift contributions, it is necessary to consider the next order in the
expansion of Eq. (\ref{9}), and the result is:
\beq
\begin{cases}
\frac{\d}{\d t}\langle\psi_1\rangle=
\beta'_0\langle\psi_1\rangle-2\beta_0\langle\psi_{1/2}^2\rangle,
\\
\frac{\d}{\d t}\langle c_1\rangle=
-\frac{1}{2}\beta'_0\langle c_1\rangle+2\beta_0c_0\langle\psi_1\rangle,
\\
\qquad\qquad +2\beta_0\langle\psi_{1/2}c_{1/2}\rangle+\beta'_0c_0\langle\psi^2_{1/2}\rangle+Df_0.
\end{cases}
\label{15}
\eeq
The lowest order contributions to diffusion and drift are therefore both $O(D)$, as they should.
Some simplifications of Eqs. (\ref{12}), (\ref{14}) and (\ref{15}), 
taking care of the singularities
of Eq. (\ref{11}) at $c=0$, are still possible and are illustrated in Appendix C.

\begin{figure}
\begin{center}
\includegraphics[draft=false,width=8cm]{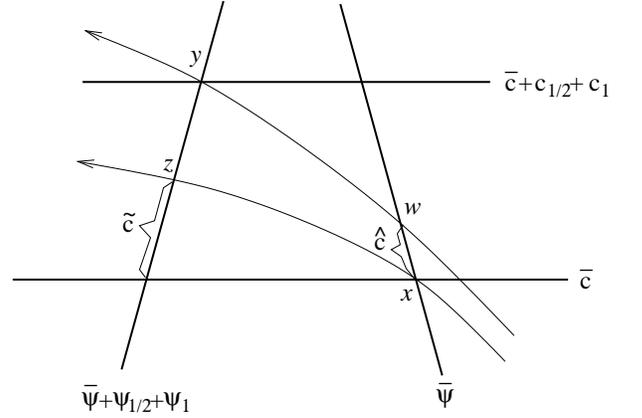}
\caption{
Orbit behavior in the proximity of the recurrent point $x=\{\bar\psi,\bar c\}$;
$x-z$ unperturbed orbit; $w-y$ noisy orbit. The deviation between orbits is identified
by $\hat c$.
}
\label{shallfig4}
\end{center}
\end{figure}

Once the noisy trajectory  
$\{\psi(t|\bar\psi),c(t|\bar c,\bar\psi)\}$
has been calculated up to the recurrence time $t_i$, it is necessary
to follow it back to the time $\hat t_i$ at which 
mod$(\psi(\hat t_i|\bar\psi),\pi)=\bar\psi$ 
and calculate the difference
$$
\hat c=c(\hat t_i|\bar c,\bar\psi)-c_0(t_i|\bar c,\bar\psi)\simeq 
c(\hat t_i|\bar c,\bar\psi)-\bar c.
$$
This operation is equivalent to the procedure, implicit in \cite{leal72}, 
of subtracting from the deviation
$\{\psi(t_i|\bar\psi)-\psi_0(t_i\bar\psi),c(t_i|\bar c,\bar\psi)-c_0(t_i|\bar c,\bar\psi)\}\simeq
\{\psi_{1/2}+\psi_1,c_{1/2}+c_1\}$,
the component along the unperturbed Jeffery's orbit, and keeping only
the part associated with percolation between the orbits. 
The necessary operations are illustrated in Fig. \ref{shallfig4}, and it is assumed
that the orbits can be parameterized locally with $\psi\equiv\psi_0(t|\bar\psi)$ 
(this is possible if $\bar\psi$ is chosen away from turning points).
The first step is to calculate the difference in $c$ between noisy and 
unperturbed orbits, at the azimuthal angle $\psi=\psi(t_i|\bar\psi)$ where
mod$(\psi(t_i|\bar\psi),\pi)\simeq \bar\psi+\psi_{1/2}+\psi_1$,
corresponding to the points $y$ and $z$ in Fig. \ref{shallfig4}.
To $O(D)$, the value of $c$ at $z$ is
\beq
\bar c+\tilde c=\bar c+c_\psi(\psi_{1/2}+\psi_1)+\frac{1}{2}c_{\psi\psi}\psi_{1/2}^2,
\label{19}
\eeq
where $c_\psi$ and $c_{\psi\psi}$ give the rise in $c$ along the unperturbed trajectory:
\beq
c_\psi=\frac{\d c_0}{\d\psi}
\quad
{\rm and}
\quad
c_{\psi\psi}=\frac{\d^2 c_0}{\d\psi^2},
\label{20}
\eeq
with $\d/\d\psi$ the derivative along the unperturbed orbit:
\beq
\frac{\d}{\d\psi}=\frac{1}{\dot\psi_0}\Big[\frac{\partial}{\partial t}+
\dot\psi_0\frac{\partial}{\partial \psi}+\dot c_0\frac{\partial}{\partial c}\Big],
\label{21}
\eeq
which is calculated at $\psi=\bar\psi$.
Combining Eqs. (\ref{20}-\ref{21}) with Eq. (\ref{12}):
\beq
\begin{cases}
c_\psi=\frac{\beta'c_0}{2(\omega-\beta_0)},
\\
c_{\psi\psi}=-\frac{\dot\beta_0\beta'_0c_0}{(\omega-\beta_0)^3}
+\frac{({\beta'_0}^2-\dot\beta'_0)c_0}{(\omega-\beta_0)^2}-\frac{2\beta_0c_0}{\omega-\beta_0},
\end{cases}
\label{21.5}
\eeq
where $\dot\beta=\partial_t\beta$ and similar for $\dot\beta'$.
To obtain $\hat c$, it is necessary to correct the difference in $c$ between $y$ and $z$, i.e.
$c_{1/2}+c_1-\tilde c$, for
the contribution from the deviation between unperturbed orbits, which, in the present case,
is $(c_{1/2}+c_1-\tilde c)(\psi_{1/2}+\psi_1)c_{\psi c}$, where 
\beq
c_{\psi c}=\partial c_\psi/\partial c=
c_0^{-1}c_\psi.
\label{21.6}
\eeq
Working again to $O(D)$:
\beq
\hat c=c_{1/2}+c_1-\tilde c- \psi_{1/2} c_{\psi c}(c_{1/2}-\tilde c),
\label{18}
\eeq
corresponding to a time along the trajectory:
\beq
\hat t_i=t_i+[\omega+\beta(\bar\psi,0)]^{-1}\psi_{1/2}+O(D).
\label{18.1}
\eeq
Using the relation $\langle\psi_{1/2}c_{1/2}\rangle=-c_0\langle\psi_1\rangle$
(see Appendix C), together with Eqs. (\ref{19}) and (\ref{18}), 
the following result for the diffusion 
and drift across Jeffery's orbits is obtained, to $O(D)$:
\beq
\begin{cases}
\langle\hat c^2\rangle=\langle c_{1/2}^2\rangle+c_\psi^2\langle\psi_{1/2}^2\rangle
-2c_\psi\langle\psi_{1/2}c_{1/2}\rangle,
\\
\langle\hat c\rangle=\langle c_1\rangle+(c_\psi c_{\psi c}-\frac{1}{2}c_{\psi\psi})
\langle\psi_{1/2}^2\rangle,
\end{cases}
\label{23}
\eeq
and, combining with Eqs. (\ref{21.5}) and (\ref{21.6}), the noise induced deviation 
between Jeffery orbits is fully determined.

\vskip 15pt 
\section{DETERMINATION OF THE ORIENTATION DISTRIBUTION}
\begin{figure}
\begin{center}
\includegraphics[draft=false,width=8cm]{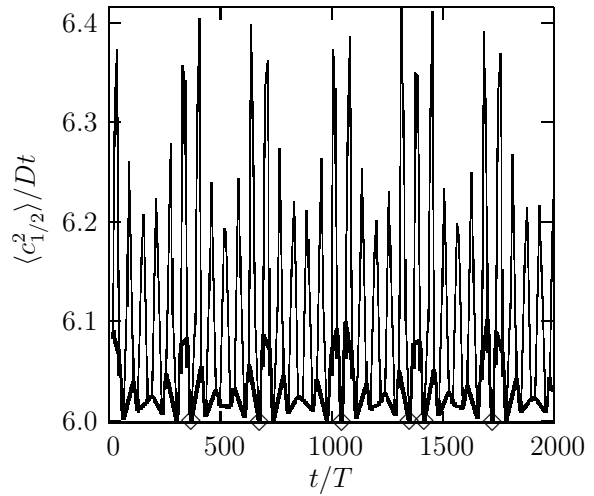}
\caption{
Determination of the normalized diffusivity $\langle c_{1/2}^2\rangle/Dt$ for different values 
of the tolerance $\epsilon$ entering the recurrence condition
$|P_n(\bar\psi)-\bar\psi|<\epsilon$ with  $\{\bar\psi,\bar c\}=\{0,0\}$.
Values of the parameters $\omega=1.4$, $\alpha\simeq 0.37$.
Thin line $\epsilon=0.4$; heavy line $\epsilon=0.1$; diamonds $\epsilon=0.01$ (the
diamonds identify the actual position of the recurrence times).
}
\label{shallfig5}
\end{center}
\end{figure}

\vskip 5pt
The quantities $\langle\hat c^2\rangle$ and $\langle\hat c\rangle$ allow to determine
the noise contribution to orbit deviation, at the corrected recurrence times 
$\hat t_i$, at which $\mod(\psi(\hat t_i|\bar\psi),\pi)=\bar\psi$.
Both quantities $\langle\hat c^2\rangle$ and $\langle\hat c\rangle$
are obtained from integrals along the orbits and
it is expected that an averaging process takes place, with 
$\langle\hat c^2\rangle/t_i$ and $\langle\hat c\rangle/t_i$
tending to finite limits as $t_i\to\infty$. 
Integrating numerically Eqs. (\ref{12}) and (\ref{14}-\ref{15}) 
[or, more simply, Eq. (C2)], with the initial condition 
$\{\psi_0(0|\bar\psi),c_0(0|\bar c,\bar\psi)\}=\{\bar\psi,\bar c\}=\{0,0\}$
and then substituting, with Eqs. (\ref{21.5}-\ref{21.6}), into 
(\ref{23}),
leads to the result in Fig. \ref{shallfig5}. Self-averaging 
of $\langle\hat c^2\rangle/t_i$ takes place
also for relatively large values of the tolerance $\epsilon$, which identifies
recurrence, and therefore the sequence $t_i=t_i(\epsilon)$ $i=1,2...$, 
through the condition  $|P_n^0(\bar\psi)-\bar\psi|<\epsilon$. 
It is thus possible to introduce 
effective drift and diffusion coefficients $\bar a$ and $\bar D$:
\beq
\bar a(\bar c,\bar\psi)=\lim_{i\to\infty}t_i^{-1}\langle\hat c\rangle,
\qquad
\bar D(\bar c,\bar\psi)=\lim_{i\to\infty}t_i^{-1}\langle\hat c^2\rangle,
\label{25}
\eeq
which will describe the dynamics of the Poincare map 
$\bar c(\hat t_i)\equiv c(\hat t_i|\bar c(0),\bar\psi)$: 
\beq
\bar c(\hat t_i)-\bar c(0)=\bar at_i+\bar D^{1/2}[W(t_i)-W(0)].
\label{25.5}
\eeq
This is a discrete Langevin equation, in which
$W(t)$ is again the Wiener process, with $\langle [W(t)-W(0)]^2\rangle=t$.
In the small $D$ limit, the recurrence times $t_i$ can be treated as continuous
on the scale of the relaxation to equilibrium. It is then possible to obtain a
Fokker-Planck equation for the evolution of the PDF for $\bar c(t)$, which,
by construction, is nothing else than $\rho(\bar c|\bar\psi)$. Now, from Eq.
(\ref{18.1}):
$$
\rho(\bar c|\bar\psi;\hat t_i)-\rho(\bar c|\bar\psi;0)=
[1+O(D^{1/2})]t_i\partial_t
\rho(\bar c|\bar\psi;t)|_{t=0},
$$
and, to lowest order in $D$, it is possible to disregard the difference between
$\hat t_i$ and $t_i$ in $\rho$; this is equivalent to substitute $\bar c(t_i)-\bar c(0)$ 
into the left hand side of Eq. (\ref{25.5}). Taking the continuous
limit, leads to the Langevin equation $\d\bar c=\bar a\d t+\bar D^{1/2}\d W$,
which is associated with the Fokker-Planck equation \cite{gardiner}: 
\beq
\partial_{\bar t}\rho+\partial_{\bar c}(\bar a\rho)=\frac{1}{2}\partial_{\bar c}^2(\bar D\rho),
\label{24}
\eeq
and the notation $\bar t$, indicating a slow time scale, is a reminder that Eq. 
(\ref{24}) is meaningful only at timescales long with respect to $t_i-t_{i-1}$.
Slow variation of the flow parameters entering Eq. (\ref{9}) would lead to dependence of
the coefficients in Eq. (\ref{24}) on the slow time $\bar t$. As in \cite{leal72},
the fact that both $\bar a$ and $\bar D$ depend linearly in $D$ implies that the
equilibrium PDF is independent of the noise amplitude.

\begin{figure}
\begin{center}
\includegraphics[draft=false,width=8cm]{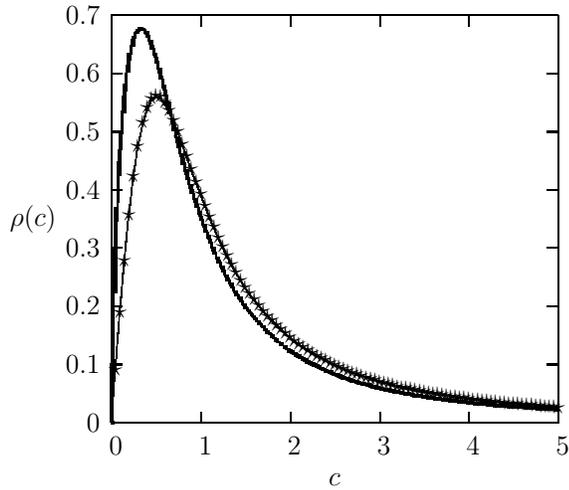}
\caption{
Comparison of the PDF $\rho(\bar c|\bar\psi)$ calculated using for $\bar a$ and $\bar D$
different values of $t_i$ [see Eq. (\ref{25})]. Values of the parameters:
$\omega=1.4$, $\alpha=0$, $\bar\psi=0$. 
Heavy line: $t_i=3T$ $\epsilon=0.4$; stars: $t_i=20T$, $\epsilon=0.1$;
thin line: Leal \& Hinch theory \cite{leal72}.
}
\label{shallfig6}
\end{center}
\end{figure}

As statistical equilibrium will be achieved on the time scale $D^{-1}$, in order
for the approach to be meaningful, it is necessary
that the $t_i$ used to define $\bar a$ and $\bar D$ 
satisfy $Dt_i\ll 1$. 
Actually, excellent convergence is obtained already for $t_i$ rather small; in the case
of Fig. \ref{shallfig6}, at $t_i=20T$, corresponding to $\epsilon=0.1$.
Notice that the case considered in Fig. \ref{shallfig6}, which is identical to 
the deep water wave regime considered
in \cite{decarolis05}, 
can be mapped to
a constant simple shear by a redefinition of the eccentricity $G$.
[In this case, the aperiodicity of $P_n^0(\bar\psi)$ 
originates not from the dynamics, but from the sampling time
$T$ and the rotation period  $T_r=(\omega^2-1)^{-\frac{1}{2}}$ being incommensurate].
The PDF
$\rho(\bar c|\bar\psi)$ can then be compared with the analytical result from the theory of Leal 
and Hinch [the function $f(C)$ in Eq.(17) of their paper] \cite{leal72}.
As can be seen from Fig. \ref{shallfig6}, the two approaches give indistinguishable results
already for $t_i=20T$, $\epsilon=0.1$. Similar convergence to the limit result is observed 
for $\alpha>0$, when comparison with the theory of Leal and Hinch is not
possible.

\begin{figure}
\begin{center}
\includegraphics[draft=false,width=8cm]{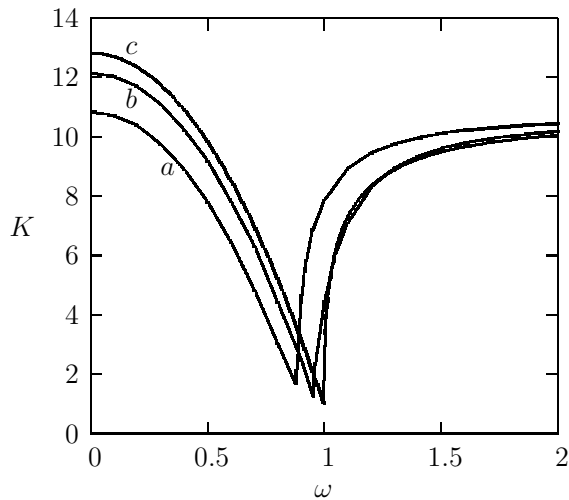}
\caption{
Reduced viscosity, averaged over a period, 
for a suspension of disk-like particles
with aspect ratio $r=0.045$. In the three cases: 
$(a)$ $\alpha\simeq 0.61$; $(b)$ $\alpha\simeq 0.37$; $(c)$ $\alpha=0$.
}
\label{shallfig7}
\end{center}
\end{figure}

Knowledge of the PDF $\rho(\bar c|\bar\psi)$ allows determination of the effective viscosity of
a dilute disk suspension in the oscillating strain field of Eq. (\ref{1}). 
The viscous stress for a suspension of
axisymmetric ellipsoids reads \cite{leal72,hinch75}, indicating with $\mu$ and $\Phi$,
respectively, the solvent viscosity and the suspended phase volume fraction:
\begin{eqnarray}
\bsigma= 
2\mu\E+2\mu\Phi 
\{2A\langle\p\p\p\p\rangle:\E
\nonumber
\\
+2B[\langle\p\p\rangle\cdot\E+
\E\cdot\langle\p\p\rangle]+C\E\},
\label{26}
\end{eqnarray}
where, in the present time dependent situation, the averages are intended
over orientation and time.
The coefficients $A-C$ depend on the particle geometry \cite{decarolis05}:
\begin{eqnarray*}
A=\frac{5}{3\pi r}+\frac{104}{9\pi^{2}}-1,
\quad
B=-\frac{4}{3\pi r}-\frac{64}{9\pi^{2}}+\frac{1}{2}
\\
{\rm and}\quad
C
=\frac{8}{3\pi r}+\frac{128}{9\pi^{2}},
\end{eqnarray*}
with $r$ the particle aspect ratio, supposed small. These expressions 
correct to $O(1)$, similar ones derived in \cite{leal72}.
From the stress $\bsigma$, the effective viscosity $\bar\mu$ can be calculated
in terms of the viscous dissipation in the suspension:
\beq
\bar\mu=\frac{1}{2}\frac{\bsigma:\E}{\E:\E}:=(1+K\Phi)\mu,
\label{28}
\eeq
where $K$ is called the reduced viscosity. 
Expressing the versor $\p$ in function of the angles $\psi$ and $\theta$,
and using Eqs. (\ref{26}) and (\ref{28}):
\beq
K=A\langle\sin^4\theta\sin^22\psi\rangle+2B\langle\sin^2\theta\rangle+C.
\label{29}
\eeq
As in \cite{leal72}, the average over orientation is split into parts along
and transverse to the orbit. In the present situation, however, evaluation
of the average along the orbit is slightly more delicate than in the
time-independent case. At a generic time $t$ the average
of a function $f(\psi,c)$ will be:
$$
\langle f\rangle(t)=\frac{1}{n}\sum_{i=1}^n\int\d\bar c\,\rho(\bar c|\bar\psi)
f(\psi(t+iT|\bar\psi),c(t+iT|\bar c,\bar\psi)),
$$
where, from ergodicity, memory of the initial condition $\bar\psi$ is lost for
$n\to\infty$.
Carrying on the average over one period, which, in the case of a wave,
from Eq. (\ref{4}), 
is equivalent also to a space average, leads to the average along an
orbit:
\beq
\langle f\rangle=\frac{1}{nT}\int\d\bar c\,\rho(\bar c|\bar\psi)\int_0^{nT}\d t\,
f(\psi(t|\bar\psi),c(t|\bar c,\bar\psi)).
\label{30}
\eeq
Evaluating Eq. (\ref{29}) with Eq. (\ref{30}) leads to the values
of the reduced viscosity shown in Fig. \ref{shallfig7}. 
The calculation has been carried on, using as recurrence point, $\bar\psi=0$;
the value 
of the particle aspect ratio has been chosen consistent with frazil ice measurements
\cite{martin81}.
The same qualitative regime observed for $\alpha=0$ is reproduced here, namely,
a dip in the reduced viscosity at the crossover from the coherent rotation regime
to the random orientation one \cite{decarolis05}.

In the case of gravity waves, the coherent regime would always be associated with 
high amplitude waves, corresponding to small values of the normalized frequency $\omega$. 
The reduced viscosity has been calculated in this range from Eq. (\ref{29}), fixing 
$\theta=\pi/2$ and integrating the equation for $\psi$, in the absence of noise, 
with initial condition at the fixed point.

\section{CONCLUSIONS}
\vskip 5pt
The numerical evaluation of the rheological properties of a suspension of particles 
that are weakly Brownian is faced with difficulties associated with the long 
integration times necessary to achieve statistical equilibrium. Analytical
techniques for the calculation of the cumulative effect of the Brownian noise 
on the dynamics are therefore necessary. The technique presented in this paper
can be seen as a multiple time scale analysis \cite{bender} in which the stochastic dynamics
is pushed to the slow scale, while the local strain and vorticity are treated as fast variables.
For the periodic flows considered in this paper, the effective drift
and diffusivity coefficients are obtained integrating the fast (and aperiodic) 
orientation dynamics up
to the first recurrence time, at which the approximation of a closed orbit 
is considered good enough.
Slow variations would be accounted for, sampling the almost closed trajectory segments
in appropriate way along the particle orbit, and would lead to effective drift
and diffusivity coefficients depending on the slow time.
Once the effective drift and diffusivity were available,
a Monte Carlo, for the determination of the rheological properties of a suspension,
would be carried on at the slow time scale [this would be associated formally with 
integration of the Fokker-Planck equation (\ref{24})].

Application of these techniques to the dynamics of a thin disk suspension in 
gravity waves, suggest that qualitative behaviors in deep water, accounted for
theoretically in \cite{decarolis05} and observed experimentally in \cite{martin81},
should be preserved in the shallow water regime.
In particular, a transition from a coherent rotation regime for 
large amplitude waves to a random orientation one, marked by a deep 
minimum in the medium effective viscosity, should continue to be present.
Away from this regime, in the random orientation regime,
the numerical values of the effective viscosity appear to be only weakly dependent 
on the water depth. This strengthens confidence in experimental data on the sea 
ice effective viscosity from wave tanks, in which the deep water condition is at 
the most only approximately satisfied, as a test case of what happens in the open 
sea \cite{martin81,newyear97}.

A natural extension of the techniques illustrated  could be 
the treatment of higher numbers of degrees of freedom. An interesting
example is the triaxial ellipsoid in a simple shear considered in 
\cite{yarin97}. In this case, the angle $\theta$ would be replaced by
the pair $\{\theta,\phi\}$ with $\phi$ the rotation around the axis $\p$.
An analysis in the whole phase domain would require, however, consideration
of the transition region from the regular orbits, in which diffusion
is dominated by Brownian rotation, to the chaos dominated stochastic 
region.  

\begin{acknowledgements}
This research was carried on at the Dipartimento
di Fisica dell'Universit\'a di Cagliari. The author whishes to thank Alberto Pompei 
for hospitality.
\end{acknowledgements}

\appendix
\section{ORIENTATION DYNAMICS IN ROTATING REFERENCE FRAME}

The strain matrix in the laboratory frame, is provided by Eq. (\ref{1}).
Passing to the rotating frame:
$$
{\bf R}\cdot{\bf E}\cdot{\bf R}^{\rm T}
=\left(
\begin{array}{rr}
1+\alpha\cos 2\omega t, & -\alpha\sin 2\omega t \\
-\alpha\sin 2\omega t, & -1-\alpha\cos 2\omega t
\end{array}
\right),
$$
where
$$
{\bf R}=
\left( 
\begin{array}{rr}
\cos\omega t/2, &  \sin\omega t/2  \\
-\sin\omega t/2, & \cos\omega t/2                
\end{array}
\right)
$$
is the matrix for an angle $-\omega t/2$ rotation.
In term of components, $u_i$ and $v_i=R_{ij}u_j$ are the components
of a vector in the laboratory and the rotating reference frame. In the rotating frame, 
the fluid is seen rotating like $v_i$:
$$
\dot v_i=\Omega_{ij}v_j,
\qquad
{\bf\Omega}=\frac{\omega}{2}\left(
\begin{array}{rr}
0, & 1\\
-1, & 0
\end{array}
\right),
$$
and $\bOmega$ is the vorticity of the fluid measured in the rotating frame.
One more rotation by $\pi/4$ produces Eq. (\ref{2})
$$
\hat{\bf R}
\left(
\begin{array}{rr}
1+\alpha\cos 2\omega t, & -\alpha\sin 2\omega t \\
-\alpha\sin 2\omega t, & -1-\alpha\cos 2\omega t
\end{array}
\right)
\hat{\bf R}^{\rm T}
$$
$$
=\left(
\begin{array}{rr}
\alpha\sin 2\omega t, & 1+\alpha\cos 2\omega t\\
1+\alpha\cos 2\omega t, &-\alpha\sin 2\omega t  
\end{array}
\right),
$$ 
where
$$
\hat{\bf R}=
\frac{1}{\sqrt{2}}
\left(
\begin{array}{rr}
1,& -1 \\
1,& 1
\end{array}
\right).
$$
As the next step, pass to adimensional variables:
$$
\hat t=-Ge t,\qquad \hat\omega=-\frac{\omega}{2Ge},\qquad \hat{\bf E}=e^{-1}{\bf E}.
$$
This choice guarantees that, in the case of oblate ellipsoids, 
the signs of the normalized times and frequency are
preserved. 
Substituting into the Jeffery's equation (\ref{5}) gives:
$$
\frac{\d\p}{\d\hat t}=\hat\omega\left(
\begin{array}{rr}
0, & 1\\
-1, & 0
\end{array}
\right)\p-[\hat\E\cdot\p - (\p\cdot\hat\E\cdot\p)\p].
\eqno(A1)
$$
Introducing polar coordinates $\p=(\sin\theta\cos\psi,\sin\theta\sin\psi,$ $\cos\theta)$ and
using Eq. (\ref{2}), leads to the expressions:
$$
\E\cdot\p=\sin\theta\left(
\begin{array}{rr}
\sin\psi+\alpha\sin(\psi+4\hat\omega t)\\
\cos\psi+\alpha\cos(\psi+4\hat\omega t)
\end{array}
\right)
\eqno(A2)
$$
and
$$
\p\cdot\E\cdot\p=\sin^2\theta[\sin 2\psi+\alpha\sin(2\psi+4\hat\omega t)].
\eqno(A3)
$$
The Jeffery's equation can now be written in components. Starting from $\theta$,
using Eqs. (A1-A3):
$$
\dot p_3=-\sin\theta\dot\theta=\cos\theta\sin^2\theta[\sin 2\psi+\alpha\sin(2\psi+4\hat\omega t)],
\eqno(A4)
$$
which leads to the second of Eqs. (\ref{6}-\ref{7}). Passing to the equation for $\psi$:
\begin{eqnarray*}
\dot p_2=\sin\theta\cos\psi\dot\psi+\cos\theta\sin\psi\dot\theta
=-\hat\omega\sin\theta\cos\psi
\\
-\sin\theta[\cos\psi+\alpha\cos(\psi+4\hat\omega t)]
+\sin^3\theta[\sin2\psi
\\
+\alpha\sin(2\psi+4\hat\omega t)]\sin\psi,
\end{eqnarray*}
from which, using Eq. (A4):
\begin{eqnarray*}
\cos\psi\dot\psi
=-\hat\omega\cos\psi-\cos\psi(1-2\sin^2\psi)
\\
+\alpha[\sin(2\psi+4\hat\omega t)\sin\psi-\cos(\psi+4\hat\omega t)],
\end{eqnarray*}
and, after little algebra:
$$
\dot\psi=
-\hat\omega-\cos 2\psi-\alpha\cos(2\psi+4\hat\omega t),
$$
which is the first of Eqs. (\ref{6}-\ref{7}).

\section{NOISE TERM DETERMINATION}
The noise term to add in Eq.  (\ref{6}) can be obtained directly
from the diffusion equation obeyed for zero flow by the orientation PDF in the variables
$\{\psi,c\}$. Alternatively, one may consider the diffusion operator in the variables
$\{\psi,\theta\}$:
$$
\nabla^2=\frac{1}{\sin^2\theta}\frac{\partial^2}{\partial\psi^2}
+\frac{1}{\sin\theta}\frac{\partial}{\partial\theta}\sin\theta\frac{\partial}{\partial\theta},
$$
and determine the stochastic process leading to the Fokker-Planck equation 
$\nabla^2\rho(\psi,\theta)=0$ 
[which has the isotropic solution $\rho(\psi,\theta)=\frac{1}{2\pi}\sin\theta$].
One finds the increments for $\psi$ and $\theta$ produced by Brownian 
rotation in the time interval $\d t$ \cite{gardiner}:
$$
\d\psi=|\sin\theta|^{-1}\d W_\psi,
\qquad
\d\theta=\frac{1}{2}\cot\theta\d t+\d W_\theta,
$$
where $\d W_k$, $k=\psi,\theta$ are the Brownian increments
$$
\langle\d W_k\rangle=0,
\qquad
\langle\d W_j\d W_k\rangle=\delta_{jk}\d t.
$$
Changing then variables from $\theta$ to $c$ and using It\^o's lemma, one finds:
\begin{eqnarray*}
\d c=\d\theta\frac{\d c}{\d \theta}+\frac{1}{2}
\langle\d W_\theta^2\rangle\frac{\d^2 c}{\d \theta^2}
\\
=\frac{1}{c}(1+c^2)(\frac{1}{2}+c^2)\d t+(1+c^2)\d W_\theta,
\end{eqnarray*}
and, using the expression $\sin\theta=c(1+c^2)^{-1/2}$ in $\d\psi$, 
Eq. (\ref{9}) is finally obtained.

\section{ALTERNATIVE FORM OF THE PERTURBED ORBIT EQUATION}
Equations (\ref{14}-\ref{15}) can be simplified, and 
the singularity in $\theta=0$ produced by the noise term in the first of
Eq. (\ref{9}) eliminated, by the change of variables:
\begin{eqnarray*}
y_1=c_0^2\langle\psi_{1/2}^2\rangle,
\quad
y_2=c_0\langle\psi_{1/2}c_{1/2}\rangle,
\\
y_3=\langle c_{1/2}^2\rangle,
\quad
y_4=c_0\langle c_1\rangle,
\quad
y_5=c_0^2\langle\psi_1\rangle.
\end{eqnarray*}
In the new variables, Eqs. (\ref{14}-\ref{15}) take the form:
$$
\begin{cases}
\dot y_1=\beta'_0y_1+D\tilde g,
\\
\dot y_2=2\beta_0y_1,
\\
\dot y_3=4\beta_0y_2-\beta'_0y_3+D\tilde h,
\\
\dot y_4=-\beta'_0y_4+2\beta_0y_2+\beta'_0y_1+2\beta_0y_5+D\tilde f,
\\
\dot y_5=-2\beta_0y_1,
\end{cases}
\eqno(C1)
$$
where, from Eq. (\ref{11}),
$\tilde g=1+c_0^2$, $\tilde h=(1+c_0^2)^2$ and $\tilde f=(1+c_0^2)(\frac{1}{2}+c_0^2)$.
Comparing the equations for $y_2$ and $y_5$, 
one sees that, thanks to the initial condition $y_k(0)=0$, $y_2=-y_5$ and then
$\langle\psi_{1/2}c_{1/2}\rangle=-c_0\langle\psi_1\rangle$; thus
the equation for $\langle\psi_1\rangle$ can be eliminated from (\ref{15}). 
Equation (C1) can then be further simplified to:
$$
\begin{cases}
\dot y_1=\beta'_0y_1+D\tilde g,
\\
\dot y_2=2\beta_0y_1,
\\
\dot y_3=4\beta_0y_2-\beta'_0y_3+D\tilde h,
\\
\dot y_4=-\beta'_0y_4+\beta'_0y_1+D\tilde f.
\end{cases}
\eqno(C2)
$$

\end{document}